\documentclass[11pt]{article}
\usepackage{xspace}
\usepackage{graphicx}
\usepackage{amsmath}
\usepackage{amssymb}
\usepackage{color}

\definecolor{Red}{rgb}{1,0,0}
\definecolor{Green}{rgb}{0,1,0}
\definecolor{Blue}{rgb}{0,0,1}
\definecolor{Black}{rgb}{0,0,0}

\def\beq{\begin{equation}}
\def\eeq#1{\label{#1}\end{equation}}
\def\eeqn{\end{equation}}

\def\beqa{\begin{eqnarray}}
\def\eeqa#1{\label{#1}\end{eqnarray}}
\def\eeqan{\end{eqnarray}}

\let\bar=\overbar

\def\Dslash{\not{\hbox{\kern-4pt $D$}}}
\def\dslash{\not{\hbox{\kern-2pt $\del$}}}

\def\msb{{\bar{\ssstyle M \kern -1pt S}}}

\def\Title#1{\begin{center} {\Large {\bf #1} } \end{center}}

\begin{document}

\Title{Probing neutrinoless double beta decay with SNO+}

\bigskip\bigskip


\begin{raggedright}  

{\it Evelina Arushanova\index{Arushanova, E.},\\
Ashley R. Back\index{Back, A. R.},\\
School of Physics and Astronomy\\
Queen Mary University of London\\
E1 4NS London, UK}\\

\end{raggedright}
\vspace{1.cm}

{\small
\begin{flushleft}
\emph{To appear in the proceedings of the Prospects in Neutrino Physics Conference, 15 -- 17 December, 2014, held at Queen Mary University of London, UK.}
\end{flushleft}
}

\section{Introduction}

One of the primary goals for SNO+ is to search for neutrinoless double beta decay ($0\nu2\beta$) in $^{130}$Te. A goal motivated by the fact that the process requires physics beyond the Standard Model, in the form of a Majorana mass term for the neutrino. To this end, the SNOLAB-based experiment, will see the 12 m diameter SNO acrylic vessel (AV) filled with nearly one kilotonne of the liquid scintillator Linear Alkyl Benzene (LAB). This will then be loaded with $^{130}$Te. At a depth of two kilometers (6000 mwe), and given its status as a class 2000 clean room, SNOLAB provides a favourable environment for a low-background experiment.

Recent interest in Majoron-emitting modes of $0\nu2\beta$ (\cite{Gando:2012pj} and \cite{Albert:2014fya}), has prompted an investigation into SNO+ sensitivity to these models. We briefly review the theory of these in Section \ref{sec-theory}.  The detailed work in modelling the expected backgrounds is summarised in  Sections \ref{sec-backgrounds} and \ref{sec-pileups}, followed by details of SNO+ sensitivity studies in Section \ref{sec-sensitivity}.

\section{Theory}\label{sec-theory}

It is generally expected that the decay be propagated by the exchange of a light Majorana neutrino, however it has been shown \cite{Schechter:1981bd} that any valid mechanism for the decay, results is a non-zero Majorana mass term for the neutrino. One class of models that could represent this mechanism, involve the emission of one or two additional scalar particles---know as Majorons. For such models, the true visible energy spectrum becomes a continuum, as opposed to the delta function expected for the standard mechanism. The visible energy spectrum is given by \cite{Bamert:1994hb}:
\begin{equation}
\frac{d\Gamma}{d\varepsilon_1d\varepsilon_2} = C(Q-\varepsilon_1-\varepsilon_2)^n\left[p_1\varepsilon_1F(\varepsilon_1)\right]\left[p_2\varepsilon_2F(\varepsilon_2)\right]
\end{equation}
where $\varepsilon_1, \varepsilon_2$ are the energies of the two emitted electrons and $p_1, p_2$ are there respective three-momenta. The spectrum is also parametrised by the Q-value ($Q$), $F$---Fermi functions for the two electrons---and an independent constant $C$. Different models for Majoron-emitting decays are largely characterised by their spectral index $n$.

\section{Backgrounds in SNO+}\label{sec-backgrounds}

\begin{figure} [H]
\centering
\includegraphics[width=0.4\columnwidth]{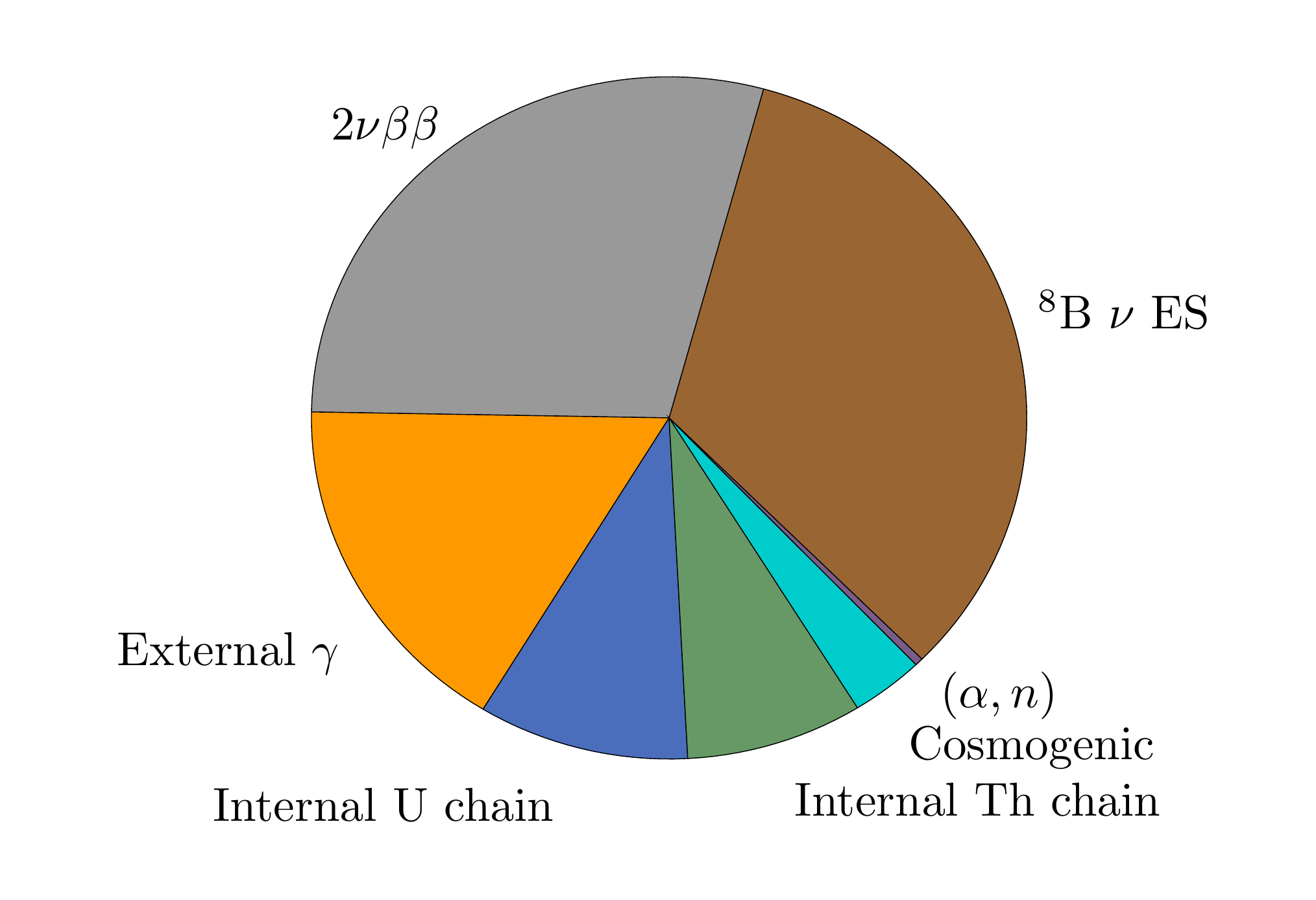}
\caption{Pie chart summarising the expected backgrounds in the ROI of 0$\nu$2$\beta$.}
\label{pie}
\end{figure}

The sources of backgrounds are shown in Fig. \ref{pie} and described below \cite{Lozza:2015}.

\textbf{${}^{8}$B and ${}^{130}$Te 2$\nu$2$\beta$}. 
Solar ${}^{8}$B and ${}^{130}$Te 2$\nu$2$\beta$ are irreducible. 

\textbf{Internal ${}^{238}$U and ${}^{232}$Th chains}.
Both backgrounds are naturally occurring isotopes inside the scintillator and reduced by careful purification process. Unless the detector is heavily contaminated these backgrounds can be managed in the analysis. 

\textbf{External gammas}.
External gammas are coming from radioactive decays that took place outside the scintillator region, such as the PMTs, PMT support structure, ropes, water shielding and the acrylic itself. Emitted alphas and majority of betas are attenuated and don't reach the scintillator. High energy gammas on the other hand can penetrate inside the LAB. External background can be reduced by applying a fiducial volume cut and a cut based on likelihood method. 

\textbf{Cosmogenic backgrounds}. 
Cosmic rays can activate the scintillator cocktail, while its components have been above ground. Luckily these background can be reduced during purification and while cooling down underground. At the depth of the SNO+ detector the cosmic rays are suppressed to an insignificant level. 

\textbf{($\alpha$, n)$\gamma$ backgrounds}. 
$\alpha$ particles, that were created in the scintillator, then can be captured by ${}^{13}$C/${}^{18}$O resulting in production of either neutrons or gammas from the excited states. The neutrons are then captured, resulting in delayed 2.2 MeV gamma emission. 

\textbf{Pileup Backgrounds}
A pileup event is a combination of multiple events, happening within the same time window of 450 ns. See Section \ref{sec-pileups} for details. 

\section{Pileup backgrounds in SNO+}
\label{sec-pileups}
\begin{figure} [H]
\centering
\includegraphics[width=0.5\columnwidth]{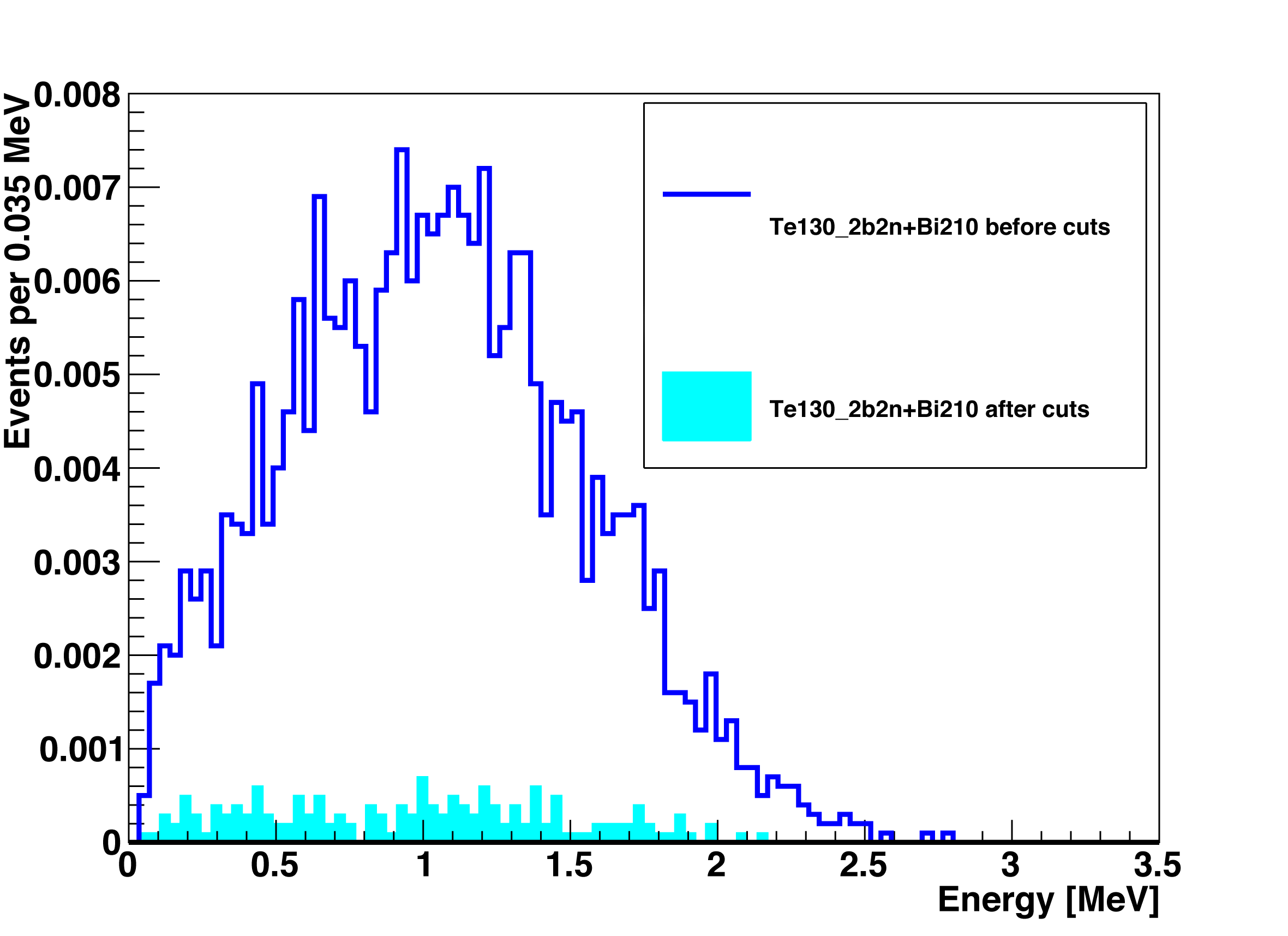}
\caption{Normalised spectra of the pileup between ${}^{130}$Te 2$\nu$2$\beta$ and ${}^{210}$Bi before and after applying the pileup rejection cuts.}
\label{pu}
\end{figure}

The end point energy of a pileup event might be higher than the end point energy of the isotopes that form it. This means that there might be additional backgrounds in the region of interest (ROI) of 0$\nu$2$\beta$. The rate of a pileup event between isotopes A and B can be described using Poisson statistics:
\begin{equation}
N_{PU} =  N_A N_B e^{-N_B}
\label{pileup}
\end{equation}
Where $N_{PU}$ is the number of the pileup events, $N_{A}$ and $N_{B}$ are the number of events of the isotopes A and B respectively. From the Eq. \ref{pileup} it can be noticed that the higher the rates of the decays, the more likely they will pileup. To be able to identify and reject the pileup events in the ROI we developed a series of algorithms and cuts, based on  the raw timing spectra. In order to demonstrate the efficiency of these techniques we  have chosen the pileup between ${}^{130}$Te 2$\nu$2$\beta$ and ${}^{210}$Bi, which is shown in Fig. \ref{pu}. The pileup backgrounds can be entirely cut in the ROI of 0$\nu$2$\beta$. 
 
\section{Sensitivity studies}\label{sec-sensitivity}

To calculate an upper-bound on the number of signal events, the number of observed events is set equal to the expected combined rate for all backgrounds described in Section \ref{sec-backgrounds} \cite{Lozza:2015}. Using a counting-experiment model, the bound can be interpreted using appropriate frequentist or Bayesian definitions of the 90\% confidence level. Before calculating the sensitivity, a 3.5 m radius fiducial volume cut is applied. Other cuts based on tagging coincidences in the $^{238}$U and $^{232}$Th chains are applied. Assuming a five year live time double beta phase of data taking, where the scintillator cocktail consists of 0.3\% $^{nat}$Te by mass, yields a projected sensitivity of  $T_{1/2}^{0\nu\beta\beta} = 9.4\times10^{25}$ yr at 90\% CL. In producing this limit the energy spectra are convolved with a Gaussian detector response function, with $\sigma = \sqrt{E [MeV] / 200 [Nhit/MeV]}$. 

\begin{figure}[!ht]\label{fig-majorons}
\begin{center}
\includegraphics[width=0.6\columnwidth]{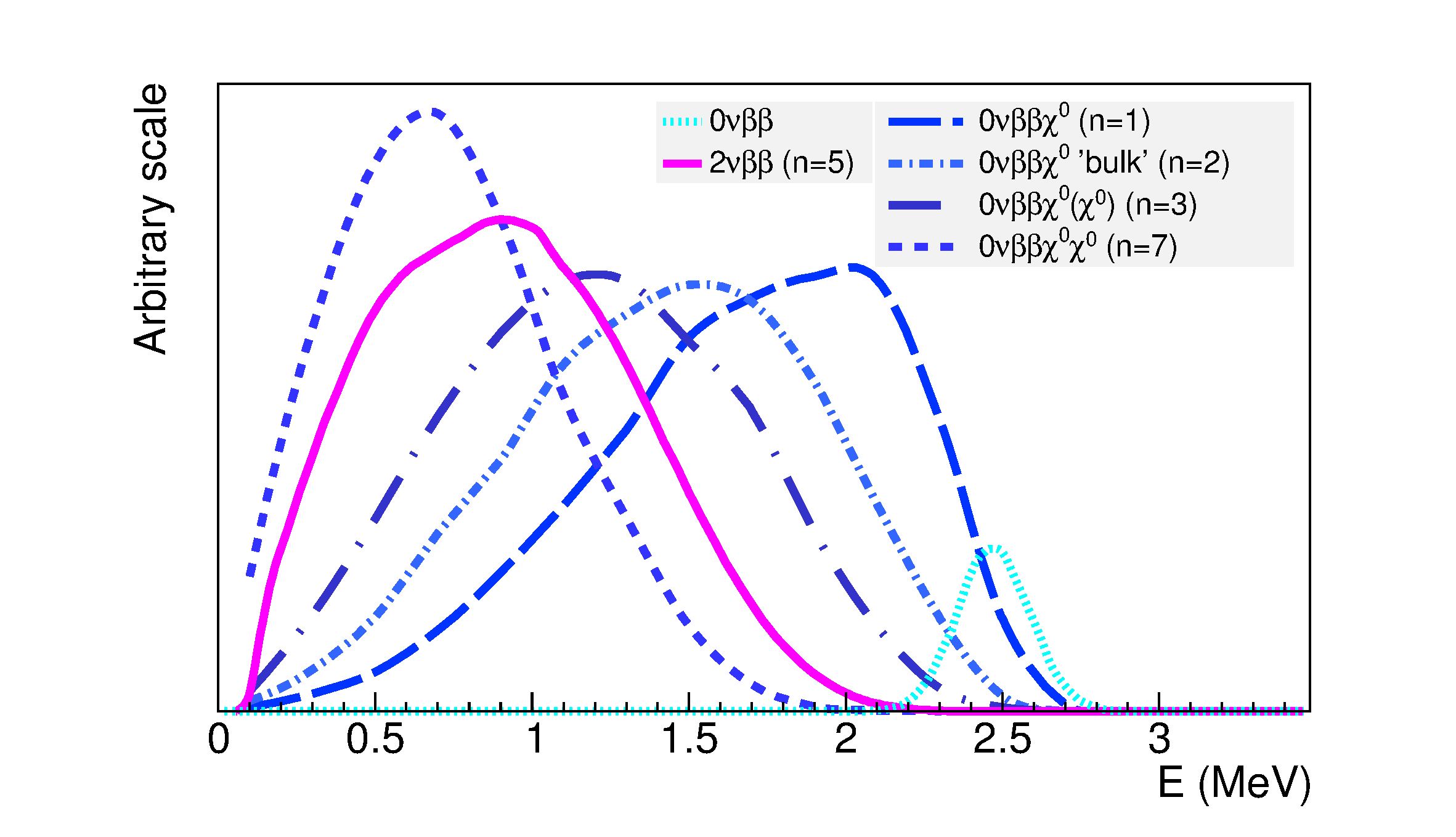}
\caption{Simulated visible energy spectra for Majoron-emitting modes ($n=1, 2, 3, 7$), all normalised to 1000 events. The $2\nu2\beta$ background contribution (also scaled to 1000 events) and a standard $0\nu2\beta$ signal (scaled to 25\% $2\nu2\beta$) are also included for reference.}
\end{center}
\end{figure}

The Majoron-emitting modes described in Section \ref{sec-theory}, provide an alternative avenue for probing $0\nu2\beta$, in addition to the above search. Examples of the visible energy spectra for Majoron-emitting models are plotted in Fig. \ref{fig-majorons}, categorised by spectral index. The continuous nature of the spectra means that sensitivity to signals of this nature need not be constrained to the ROI used for the standard $0\nu2\beta$ search. An in-depth study is currently underway for SNO+ sensitivity, but since $2\nu2\beta$ remains the dominant background for these modes, it is expected that estimated SNO+ sensitivities will be competitive.

\section{Summary}
Since the pileup rejection techniques have proven to be effective in ROI of 0$\nu$2$\beta$, we are planning to develop them for the low energy region for Majoron studies. We also plan to produce a detailed sensitivity study for the Majoron-emitting modes introduced here.

\bigskip
\section{Acknowledgments}
This work has received funding from the NExT Institute and the European Research Council under the European Union's Seventh Framework Programme (FP/2007-2013)/ERC Grant Agreement no. 278310.

\end{document}